\documentclass[conference,a4paper]{IEEEtran}

\usepackage{graphicx,cite,epsfig,amssymb,amsmath}
\usepackage{pifont}
\usepackage{stmaryrd}
\usepackage{bbding}
\usepackage{subfigure}
\usepackage{algorithm}
\usepackage{algorithmic}
\usepackage{slashbox,multirow}
\usepackage{mathrsfs}
\usepackage{latexsym}
\usepackage{indentfirst}
\usepackage{fmtcount}
\usepackage{cite}
\usepackage{url}

\usepackage{bigstrut,multirow,rotating}
\newcommand{\tabincell}[2]{\begin{tabular}{@{}#1@{}}#2\end{tabular}}

\allowdisplaybreaks[4]

\IEEEoverridecommandlockouts

\begin{document}
\title{Mobile Conductance in Sparse Networks and Mobility-Connectivity Tradeoff}

\author{\authorblockN{Huazi~Zhang$^{\dag\ddag}$, Yufan~Huang$^{\ddag}$, Zhaoyang~Zhang$^{\dag}$, Huaiyu~Dai$^{\ddag}$
\\$\dag$. Dept. of Information Science and Electronic Engineering, Zhejiang University, China.
\\$\ddag$. Department of Electrical and Computer Engineering, North Carolina State University, USA
\\Email: \{hzhang17, yhuang20, huaiyu\_dai\}@ncsu.edu, ning\_ming@zju.edu.cn}}
\maketitle

\begin{abstract}
In this paper, our recently proposed mobile-conductance based analytical framework is extended to the sparse settings, thus offering a unified tool for analyzing information spreading in mobile networks. A penalty factor is identified for information spreading in sparse networks as compared to the connected scenario, which is then intuitively interpreted and verified by simulations. With the analytical results obtained, the mobility-connectivity tradeoff is quantitatively analyzed to determine how much mobility may be exploited to make up for network connectivity deficiency.
\end{abstract}

\section{Introduction}
\subsection{Motivation and Related Works}
\noindent In many emerging large-scale networks, such as sensor networks, vehicular ad hoc networks and social networks, an important application is to spread the information quickly and efficiently over the network. Information dissemination in \emph{static} networks has been well studied in literature (e.g. \cite{Gossip}). There are also many existing works on routing in mobile networks, with emphasis on protocol and algorithm development. Recently some interesting analytical results for information spreading in dynamic wireless networks have started to emerge (see \cite{ZKongMobiHoc2008,PJacquetTIT2010,Spreading-MEG,AClementiICALP2009,Pettarin2011,Gossip-mobility,Sunlei2013MobileCRN,Mobile-conductance,Mobile-conductance-TWC} and references therein). However, it is observed that most existing analytical works focus on specific mobility models, in particular random-walk like mobility models. In our recent work \cite{Mobile-conductance}, a general analytical framework is proposed for information spreading in mobile networks, which is based on a newly proposed performance metric, mobile conductance. Our analytical framework allows separation of the mobility details from the study of mobile spreading time, so it can address various types of mobility models.

In \cite{Mobile-conductance}, some relaxation on the the node transmission range $r$ was also made. Instead of assuming an always-connected network, we only required the network remains connected under mobility (more concrete discussion will be given below). Nonetheless, such an assumption is still a limitation and hinders our study in sparse networks. In this work, we endeavor to extend our analytical framework to a general choice of $r$, so long as the expected meeting time between the message set and the non-message set is finite. Note that disconnected networks are widely seen in networks with sparse population and/or with secondary licence \cite{Sunlei2013CRBlackhole,Xiaoming2014ComST} and/or under malicious attack\cite{Li2013GC}. Our results conform with existing analysis in the sparse regime (e.g. \cite{AClementiICALP2009}), yet assume more generality and wider applicability.

\subsection{Summary of Contributions}
\begin{enumerate}

\item We extend the evaluation of mobile conductance to a general scenario for $r$, thus present a unified analytical framework for information spreading in mobile networks that can accommodate various choices of mobility patterns, moving speed, and transmission range.

\item Closed-form analytical results are obtained for mobile conductance of some popular mobility models in sparse networks (when $n{r^2} = o\left( 1 \right)$ for a network of size $n$), exhibiting an interesting $\Theta (n r^2)$ penalty factor as compared to their counterpart in connected\footnote{Actually a weaker condition $r=\Omega(\sqrt{1/n})$ would suffice.} networks. This performance gap is further justified with some intuitive explanation and network simulations.

\item A quantitative tradeoff analysis between mobility and connectivity in terms of information spreading effectiveness is given, which provides insights into how mobility may be exploited to compensate for network connectivity deficiency.
\end{enumerate}

\section{Problem Formulation}
\subsection{Network and Mobility Model}
\noindent We briefly introduce the system model for completeness, and more details can be found in \cite{Mobile-conductance}. Consider an $n$-node mobile network on a unit square $\Psi$, modeled as a discrete-time Markovian evolving graph $G_{t}\triangleq(V,E_t)$, where $V \triangleq [n]$ is the vertex set and $E_t, t \in \mathbb{N}$ is the time-varying edge set. The position of node $i$ at time $t$ is denoted by $X_i(t)$. The speed of node $i$ at time $t$ is defined by $v_i(t)=|X_i(t+1)-X_i(t)|$, assumed upper bounded by $v_{max}$ for all $i$ and $t$. A common transmission range $r$ is assumed for all nodes, and two nodes are neighbors if they are within distance $r$ at some time instant.

The moving processes of all nodes $\{X_i(t), t \in \mathbb{N}\}$, $i \in [n]$, are assumed to be independent stationary Markov chains, each starting from its stationary distribution with the transition distribution $q_i$, and collectively denoted by $\{\mathbf{X}(t), t \in \mathbb{N}\}$ with the joint transition distribution $Q=\prod\limits_{i = 1}^n {{q_i}}$. Our model allows general forms of $\{{q_i}\}$; in this paper, however, we will focus on several popular mobility models further detailed below.


\subsection{Gossip-based Mobile Spreading}
\noindent Without loss of generality, we consider the problem of single-piece information dissemination through a natural randomized gossip algorithm \cite{Gossip}, and adopt the ``Move-and-Gossip" paradigm first proposed in \cite{Mobile-conductance} to facilitate the analysis. Specifically, each time slot is divided into two phases: each node first \emph{moves} and then \emph{gossips} with \emph{one} of its \emph{new} neighboring nodes.

Denote by $S\left( t \right)$ the set of informed nodes (with $S\left( 0 \right) = \{s\}$), at the \emph{beginning} of time slot $t$. Note that the node position $X_i(t)$ changes in the middle of each time slot (after the move step), while $S(t)$ is updated at the end (after the gossip step). $P_{ij}(t+1)$ is used to denote the the probability that node $i$ contacts one of its \emph{new} neighbors $j \in {\cal N}_i(t+1)$ in the gossip step of slot $t$, set as $1/|{\cal N}_i(t+1)|$ for $j \in {\cal N}_i(t+1)$, and $0$ otherwise. The metric of interest is the $\varepsilon$-spreading time, defined as:
\begin{align}\label{spreading-time}
 T_{\mathrm{spr}} \left( \epsilon  \right)
 \triangleq  {\sup }_{s \in
V} \inf \left\{ {t:\Pr \left( {\left|S\left( t \right)\right| \ne n\left|
{S\left( 0 \right) = \left\{ s \right\}} \right.} \right) \le
\epsilon } \right\}.
\end{align}

%
%

Mobile conductance is defined for a stationary Markovian evolving graph as
\begin{equation}\label{mobile-conductance}
 {\Phi _m}\left( Q \right)
\triangleq \mathop {\min }\limits_{\scriptstyle {S'(t) \subset V}\atop
\scriptstyle {\left| {S'\left( t \right)} \right| \le n/2} }   \left\{  \mathbb{E}_Q \left( {\frac{{\sum\limits_{ i \in S'\left( t \right), j \in \overline {S'\left( t \right)} } {{P_{ij}}\left( {t + 1}\right)} }}{{\left| {S'\left( t \right)} \right|}}} \right) \right\}.
\end{equation}
Some explanations are in order: (i) $\{P_{ij}(t+1)\}$ depend on the network topology after the move, so should be considered as random variables; (ii) thanks to the stationary Markovian assumption, their expected values are well defined with respect to the transition distribution $Q$; (iii) minimization over the choice of $S'(t)$ essentially determines the bottleneck of information flow in the mobile setting.

A careful examination of our derivation of $T_{spr}$ in \cite{Mobile-conductance} reveals that no connectivity requirement is imposed. Thus,
\begin{equation}\label{mobile-spreading-time}
T_{spr} \left( \varepsilon, Q  \right) = O\left( {\frac{{\log n + \log
\varepsilon ^{ - 1} }}{{\Phi _m(Q) }}} \right)
\end{equation}
should hold for mobile spreading time in a general setting. In this work, we endeavor to evaluate \eqref{mobile-conductance} for a general $r$. For concreteness, we assume the celebrated random geometric graph (RGG) model \cite{Gossip} for the initial node distributions, i.e., $G_0 = G(n,r)$.

\section{Re-evaluation of Mobile Conductance}\label{Sect.Mobility.Model}
\noindent In our previous work \cite{Mobile-conductance}, it is assumed that the network remains connected under mobility, i.e., at each instant there exist some contact pairs between the message set and non-message set after the move. Mathematically, this means $\mathbb{E}_Q[N_{S'}(t+1)]>0$, where ${N_{S'}}\left( t+1 \right)= \sum_{i \in S'\left( t \right),j \in \overline {S'\left( t \right)} } {{1_{ij}}\left( j \in {\cal N}_i(t+1) \right)}$\footnote{In this section, $S'\left( t \right)$ intuitively takes the role of the message set, conforming to the definition in \eqref{mobile-conductance} }. Under such an assumption, the contact probabilities $P_{ij}(t+1)$'s are on the order of $P(n,r)=\Theta\left(\frac{1}{{n\pi {r^2}}}\right)$ for many popular random-walk based mobility models (including what we discuss below) in RGG, and mobile conductance admits a simpler expression \cite{Mobile-conductance}.  In sparsely populated networks, this fundamental assumption no longer holds and the previous method fails. Hence, we develop a new method to directly evaluate mobile conductance in \eqref{mobile-conductance}, which works for a general $r$; the results we obtain with this method agree with those in \cite{Mobile-conductance} and reveals a penalty factor of $\Theta (n r^2)$ when $n{r^2} = o\left( 1 \right)$.

\begin{table}
  \centering
  \caption{Highlights of Key Differences}
    \begin{tabular}{|c|c|c|}
    \hline
          & Number of Contact Pairs & Contact Probability   \bigstrut\\
    \hline
    \hline
    \multicolumn{1}{|c|}{\multirow{3}[4]{*}{\tabincell{c}{Connected\\ Graph}}} & \multicolumn{1}{c|}{\multirow{2}[2]{*}{$|{\cal N}_i(t+1)|\approx n\pi {r^2} \gg 1$}} & \multicolumn{1}{c|}{\multirow{2}[2]{*}{$P_{ij}(t+1)\approx \frac 1 {n\pi {r^2}}$}} \bigstrut[t]\\
    \multicolumn{1}{|c|}{} & \multicolumn{1}{c|}{} & \multicolumn{1}{c|}{} \bigstrut[b]\\
    \cline{2-3}
    \multicolumn{1}{|c|}{} & \multicolumn{2}{c|}{ ${\Phi _m(Q)} \propto P\left( {n,r} \right)\mathbb{E}_Q\left[{N_{S'}}\left( {t + 1} \right)\right]$ \quad \cite{Mobile-conductance} } \bigstrut\\
    \hline
    \multicolumn{1}{|c|}{Remark} & \multicolumn{2}{c|}{Indirectly calculating the number of contact pairs.} \bigstrut\\
    \hline
    \hline
    \multicolumn{1}{|c|}{\multirow{3}[4]{*}{\tabincell{c}{\\ General\\ Graph}}} & \multicolumn{1}{c|}{\multirow{2}[2]{*}{$\left| {{{ {\cal N}}_i}\left( {t + 1} \right)} \right|=\{0,1,...\}$}} & \multicolumn{1}{c|}{\multirow{2}[2]{*}{$P_{ij}(t+1) = \frac 1 {|{\cal N}_i(t+1)|}$}} \bigstrut[t]\\
    \multicolumn{1}{|c|}{} & \multicolumn{1}{c|}{} & \multicolumn{1}{c|}{} \bigstrut[b]\\
    \cline{2-3}
    \multicolumn{1}{|c|}{\multirow{2}[2]{*}{\tabincell{c}{}}} & \multicolumn{2}{c|}{${\Phi _m(Q)} \propto {\mathbb{E}_Q}\left(\sum\limits_{i \in S'(t), j \in \overline {S'(t)}} {{P_{ij}}\left( {t + 1} \right)}\right)$} \bigstrut\\
    \hline
    \multicolumn{1}{|c|}{Remark} & \multicolumn{2}{c|}{Directly calculating the sum of contact probabilities.} \bigstrut\\
    \hline
    \end{tabular}%
  \label{tab:KeyDifferences}%
\end{table}%

The key differences between the connected and general cases are highlighted in Table \ref{tab:KeyDifferences} and will be illustrated through the derivation for the fully random mobility model in III.A. In the interest of space, we simplify the derivations for three other mobility models and refer the reader to \cite{Mobile-conductance-general-R} for details. Besides, we characterize the ``mobility intensity" for each of these three mobility models with respect to the fully random mobility model, which will facilitate our mobility-connectivity tradeoff analysis in the following section.
%

\subsection{Fully Random Mobility}
\noindent \emph{Definition \cite{Mobility-throughput}:} $X_i(t)$ is uniformly distributed on $\Psi$ for all $i \in V$, i.i.d. over time. This idealistic model is adopted to explore the largest possible performance improvement brought about by mobility.

As shown in Table \ref{tab:KeyDifferences}, the main challenge lies in directly calculating the expected sum of contact probabilities. For this mobility model, the probability that an arbitrary node $j$ belongs to ${\cal N}_i(t+1)$, $\forall i \in S'\left( t \right)$ is given by
\begin{equation}\label{p_i-j}
{p_{i \leftrightarrow j}} = \pi {r^2}.
\end{equation}
The probability that $i$ has $m$ neighbors is
\begin{align}\label{P_i,m}
{p_{i,m}} &\triangleq \Pr \left\{ {\left| {{{ {\cal N}}_i}\left( {t + 1} \right)} \right| = m} \right\} \nonumber \\
&= {n-1 \choose m}{\left( {{p_{i \leftrightarrow j}}} \right)^m}{\left( {1 - {p_{i \leftrightarrow j}}} \right)^{n - 1 - m}}.
\end{align}
Among the $m$ neighbors, the probability that $b$ of them comes from $\overline {S'\left( t \right)}$ is
\begin{equation}\label{P_m,b}
p_{m,b} = {m \choose b}{\left( {{p_{i, \overline {S'}}}} \right)^b}{\left( {1 - {p_{\overline {i, S'}}}} \right)^{m-b}},
\end{equation}
where ${p_{i, \overline {S'}}}$ denotes the probability of meaningful-contact, in which a uniformly and randomly chosen edge connects $i \in {S'\left( t \right)}$ with another node from $\overline {S'\left( t \right)}$ and thus offers an effective information transfer. Thanks to the uniform node distribution of ${S'\left( t \right)}$ and $\overline {S'\left( t \right)}$ after the move, $p_{i, \overline {S'}}$ is identical for all $i$'s and given by ${p_{i, \overline {S'}}}  = \left|\overline {S'\left( t \right)}\right| / (n-1)$.

With the gossip constraint, the expected sum of contact probabilities related to node $i \in {S'\left( t \right)}$ is given by
\begin{equation}
{{\mathbb E}_Q}\left( {\sum\limits_{j \in \overline {S'\left( t \right)} } {{P_{ij}}\left( {t + 1} \right)} } \right) = \sum\limits_{m = 1}^{n - 1} {{p_{i,m}}\sum\limits_{b = 1}^m {\left( {\frac{b}{m}{p_{m,b}}} \right)} }.
\end{equation}

By the uniformity of all nodes in ${S'\left( t \right)}$, the quantity of our interest may be evaluated as
\begin{align}\label{mc2}
&{{\mathbb E}_Q}\left( {\sum\limits_{i \in S'\left( t \right),j \in \overline {S'\left( t \right)} } {{P_{ij}}\left( {t + 1} \right)} } \right) \nonumber \\
= & \left| {S'\left( t \right)} \right|{{\mathbb E}_Q}\left( {\sum\limits_{j \in \overline {S'\left( t \right)} } {{P_{ij}}\left( {t + 1} \right)} } \right) \nonumber \\
= & \left| {S'\left( t \right)} \right|\sum\limits_{m = 1}^{n - 1} {\frac{1}{m}} {p_{i,m}}\sum\limits_{b = 1}^m {b{p_{m,b}}} \nonumber  \\
= & \left| {S'\left( t \right)} \right|\sum\limits_{m = 1}^{n - 1} {\frac{1}{m}} {p_{i,m}}m\frac{{\left| {\overline {S'\left( t \right)} } \right|}}{n-1} \nonumber  \\
= & \frac{{\left| {S'\left( t \right)} \right|\left| {\overline {S'\left( t \right)} } \right|}}{n-1}\left( {1 - {{\left( {1 - {p_{i \leftrightarrow j}}} \right)}^{n - 1}}} \right)
\end{align}

Combining \eqref{p_i-j}, \eqref{mc2} and \eqref{mobile-conductance}, we have
\begin{align}
 {\Phi^{FR} _m}\left( Q \right)
&= \mathop {\min }\limits_{\scriptstyle {S'(t) \subset V}\atop
\scriptstyle {\left| {S'\left( t \right)} \right| \le n/2} } \left\{\frac{{\left| {\overline {S'\left( t \right)} } \right|}}{n-1} \left({1 - {{\left( {1 - \pi {r^2}} \right)}^{n-1}}}\right) \right\} \nonumber \\
& \simeq \frac 1 2 \left(1 - {\left( {1 - \pi {r^2}} \right)^{n-1}} \right) \label{mc-fr-general} \\
& = \left\{ \begin{matrix}
   {\Theta \left( {n{r^2}} \right),} & {n{r^2} = o(1),}\\
   {\Theta \left( 1 \right),} & {n{r^2} = \Omega(1).} \end{matrix}  \right.  \label{mc-fr-general-order}
\end{align}

As can be seen from \eqref{mc-fr-general-order}, for sparse networks with $r=o(\sqrt{1/n})$, information spreading can still be achieved (thanks to the high node mobility), but with a penalty factor of $\Theta (n {r^2})$ when compared to the connected scenario.

\subsection{Velocity Constrained Mobility}
\noindent \emph{Definition \cite{Velocity-mobility}:} This is one generalization of the fully random mobility model, with all node speeds bounded by an arbitrary $v_{\max}=O(1)$. For this model, the maximum node velocity $v_{\max}$ is considered as the mobility intensity as compared to the fully random mobility model.

\begin{figure}[h] \centering
\includegraphics[width=0.5\textwidth]{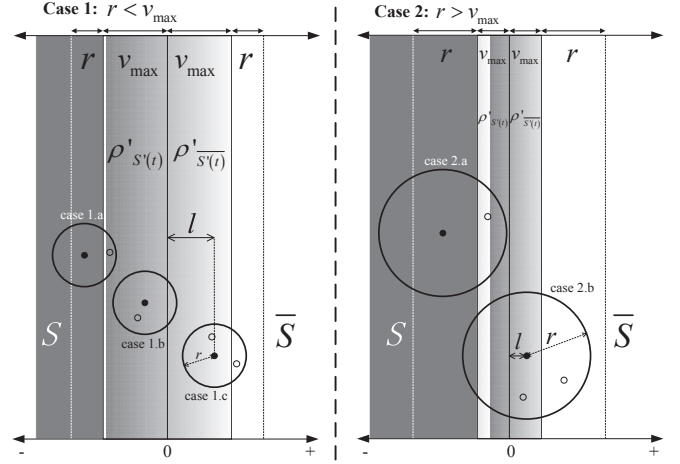}
\caption {Calculation of the $\{l,r,v_{\max}\}$ meaningful-contact probability}
\label{General-Contact-Probability-Case-1AND2}
\end{figure}

It can be shown that the bottleneck segmentation (i.e., the cut that achieves the minimum in \eqref{mobile-conductance}) remains the same as in \cite{Mobile-conductance}, and a bisection of the unit square with $S'(t)$ on the left half-plane serves this purpose. Given this setting, each node may move uniformly to any point within the circle of radius $v_{\max}$ centered at its original position. The node distribution of $S'(t)$ and $\overline {S'(t)}$ after the move has been well studied in \cite{Mobile-conductance}, and the overall diffusion process may be illustrated in Fig.~\ref{General-Contact-Probability-Case-1AND2}, in which the darkness level of the area represents the density of nodes belonging to $S'\left( t \right)$.

The key differences with the fully random mobility is that, $p_{i, \overline {S'}}$, the probability of meaningful contact (see description after Equation \eqref{P_m,b}), is different for different $i$'s, since the distribution of nodes in $S'(t)$ and $\overline {S'(t)}$ after the move is no longer uniform. Instead, $p_{i, \overline {S'}}$ depends on node $i$'s position, transmission radius $r$ and node speed $v_{\max}$, formally defined as follows.

\emph{Definition:} the $\{l,r,v_{\max}\}$ meaningful-contact probability, ${p_{i, \overline {S'}}}\left( {l,r,v_{\max}} \right)$, is the probability that, under transmission radius $r$ and velocity constraint $v_{\max}$, a uniformly and randomly chosen edge of node $i \in {S'\left( t \right)}$ with $X$-coordinate $l$ connects $i$ with another node in $\overline {S'\left( t \right)}$.

The technical challenges we face are two-fold. First, we need to calculate ${p_{i, \overline {S'}}}\left( {l,r,v_{\max}} \right)$ for each position $l$ under non-uniform distribution. Second, calculating the expected sum of contact probabilities (similar to the processes in \eqref{p_i-j} -- \eqref{mc2}) using ${p_{i, \overline {S'}}}\left( {l,r,v_{\max}} \right)$ becomes mathematically involved.

Intuitively, ${p_{i, \overline {S'}}}\left( {l,r,v_{\max}} \right)$ corresponds to the proportion of nodes without message among all $i$'s neighbors (the $r$-radius circle centered at $i$'s position \emph{after the move}). Rigorously, ${p_{i, \overline {S'}}}\left( {l,r,v_{\max}} \right)$ may be calculated by integrating the density of nodes belonging to $\overline {S'(t)}$ within the circle, and takes positive value only when $-v_{\max}-r < l < v_{\max}$. As seen in Fig.~\ref{General-Contact-Probability-Case-1AND2}, it will be a piecewise integral that involves three or two segments, depending on whether $r<v_{\max}$ and $r>v_{\max}$, denoted by \emph{case 1} and \emph{case 2}, respectively.

Following a similar approach as in III.A, but with much more tedious math, we can obtain the following results. For sparse networks ($n r^2 = o(1)$), the mobile conductance scales as $\Theta \left( {n{v_{\max }}{r^2} + n{r^3} + \frac{{n{r^4}}}{{{v_{\max }}}}} \right)$ for case 1, and as $\Theta \left( {n{v_{\max }}{r^2} + n{r^3} + nv_{\max }^3} \right)$ for case 2. After some simplification, we have
\begin{equation}\label{general-velocity-conduct-simplified}
{\Phi^{VC} _m}\left( Q \right) = \left\{ {\begin{array}{*{20}{c}}
   {\Theta \left( n{r^2} \max \left( v_{\max },r \right) \right),  \ {n{r^2} = o\left( 1 \right)},}  \\
   {\Theta \left(  \max \left( v_{\max },r \right) \right),  \ {n{r^2} = \Omega \left( 1 \right)}.}  \\
\end{array}} \right.
\end{equation}
\eqref{general-velocity-conduct-simplified} \textit{unifies} the results in \cite{Spreading-MEG} (for connected networks) and \cite{AClementiICALP2009} (for disconnected networks), and extends the study to the general $r$ scenario (c.f. $r=\Omega(\sqrt{1/n})$ in \cite{AClementiICALP2009}). The $\Theta(n r^2)$ penalty is again observed.


\subsection{Partially Random Mobility}
\noindent \emph{Definition \cite{Mobile-conductance}:} $k$ randomly pre-selected nodes are mobile, following the fully random mobility model, while the rest $n-k$ nodes stay static. This is another generalization of the fully random mobility model. For this model we consider $\frac k n$, the proportion of mobile nodes in the network, as an indicator for mobility intensity, coined as the ``mobility ratio".

Following a similar approach as above, we can obtain the following result:
\begin{align}\label{general-partial-conduct-simplified}
&{\Phi^{PR} _m}\left( Q \right) = \nonumber \\
&\left\{ {\begin{array}{*{20}{c}}
   {{{\left( {\frac{{n - k}}{n}} \right)}^2}{\Theta\left(n{r^3}\right)} + \Theta \left( {\frac{{k\left( {2n - k} \right)}}{n}{r^2}} \right),} {\ n{r^2} = o\left( 1 \right),}  \\
   {{{\left( {\frac{{n - k}}{n}} \right)}^2}{\Theta\left(r\right)} + \Theta \left( {\frac{{k\left( {2n - k} \right)}}{{{n^2}}}} \right),} {\ n{r^2} = \Omega \left( 1 \right).}  \\
\end{array}} \right.
\end{align}
For both sparse and connected scenarios, the mobile conductance in \eqref{general-partial-conduct-simplified} comprises two components, the former corresponding to static nodes while the latter corresponding to mobile nodes; a $\Theta(n r^2)$ gap is again observed between the two scenarios.

\subsection{One-Dimensional Mobility}
\noindent \emph{Definition \cite{One-dim-mobility-1}:} In this model among the $n$ nodes, $n_V$ nodes only move vertically (V-nodes) and $n_H$ nodes only move horizontally (H-nodes). It is assumed that both V-nodes and H-nodes are uniformly and randomly distributed on $\Psi$, and the the mobility pattern of each node is ``fully random" on the corresponding one-dimensional path. For this model we consider $\frac {n_V n_H} {n^2}$ as the ``mobility balance" to represent the degree of polarization in the nodes' moving directions.

The following result can be obtained after some work:
\begin{align}\label{general-1d-conduct-simplified}
{\Phi^{1D} _m}\left( Q \right) = \left\{ {\begin{array}{*{20}{c}}
   {\frac{{n_V^2 + n_H^2}}{{{n^2}}}{\Theta\left(n{r^3}\right)} + \Theta \left( {\frac{{{n_V}{n_H}}}{n}{r^2}} \right),} {\ n{r^2} = o\left( 1 \right),}  \\
   {\frac{{n_V^2 + n_H^2}}{{{n^2}}}{\Theta\left(r\right)} + \Theta \left( {\frac{{{n_V}{n_H}}}{{{n^2}}}} \right),} {\ n{r^2} = \Omega \left( 1 \right),}  \\
\end{array}} \right.
\end{align}
from which a $\Theta(n r^2)$ gap between the sparse and connected scenarios is again observed.

\section{Verification of the $n r^2$ Gap}
\subsection{An Intuitive Explanation}
\noindent The $\Theta(n r^2)$ gap between the sparse and connected networks is observed for all the above four mobility models, which is not evident from the theoretical analysis. Here we offer an intuitive explanation, relating it to the ``edge use ratio", coined by us to indicate the proportion of available edges between a node and its neighbors actually used for message spreading.

\begin{figure}[h] \centering
\includegraphics[width=0.5\textwidth]{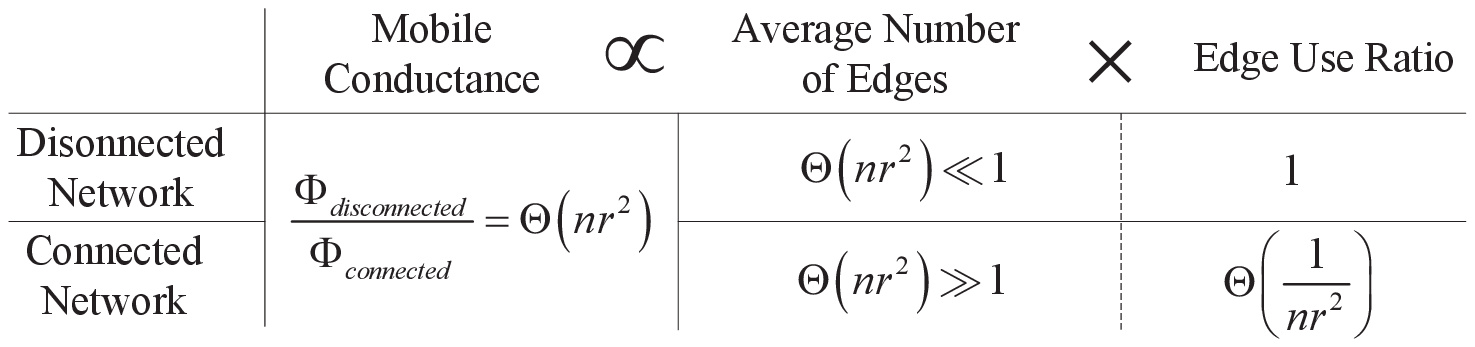}
\caption {Intuition on the $n r^2$ Gap}
\label{Gap-Intuition}
\end{figure}

On the bottommost level, the conductance is determined by the sum of contact probabilities, rather than the sum of contact pairs. Roughly speaking, given the same bottleneck segmentation, it is proportional to the product of the average number of neighbors per node and the average edge use ratio.
A comparison of the two extreme cases is given in Fig.~\ref{Gap-Intuition}.

As can be seen, in fully connected networks, a node usually has many neighbors, but can only contact one of them with the gossip constraint. In contrast, in extremely sparse networks, a node may have no neighbors with high probability; but in the rare case that it does have a neighbor, the edge use ratio is 100\%.  With this intuition in mind, the results obtained in our derivation may be better understood.

\subsection{Simulation Results}
\begin{figure}[h] \centering
\includegraphics[width=0.45\textwidth]{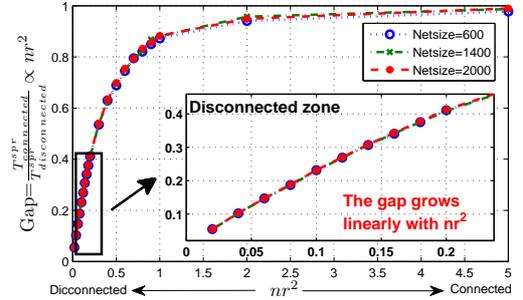}
\caption {The gap v.s. $nr^2$ in fully random mobility model}
\label{Fullrandom-Gap}
\end{figure}

\begin{figure}[h] \centering
\includegraphics[width=0.45\textwidth]{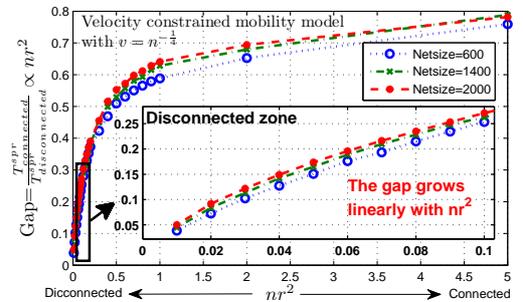}
\caption {The gap v.s. $nr^2$ in velocity constrained mobility model}
\label{Velocity-Gap}
\end{figure}

\noindent We further confirm this performance gap through simulations. Since mobile spreading time is inversely proportional to mobile conductance, the gap can be observed through the ratio of the spreading time in the fully connected and sparse networks. The results for the fully random mobility and velocity constrained mobility models are given in Fig.~\ref{Fullrandom-Gap} and Fig.~\ref{Velocity-Gap}.

%

In both figures, we can see that the gap grows roughly linearly with $nr^2$ in the disconnected zone and becomes saturated in the connected zone.
The $\Theta (n r^2)$ gap turns out to be quite accurate even for not-so-large network sizes.

\section{Mobility-Connectivity Tradeoff for Information Spreading}\label{dynamism-connectivity-tradeoff}
\noindent One striking observation from above study is that, even in a network with transmission range $r$ far below the connectivity threshold, information can still be spread granted sufficient mobility. A question naturally arises as to how much mobility is needed to facilitate information spreading given a certain degree of connectivity. Aided by our mobile conductance evaluations, some quantitative tradeoff between network dynamism and connectivity is revealed below, where we focus on a disconnected network with $n r^2 = o(1)$ (or $r=o(1/\sqrt{n})$).

The performance benchmark we consider is the information spreading time on a static ring graph. As indicated in \cite{Gossip}, the ring graph is essentially the most constrained graph in communications, and its gossip time, $\Theta \left( {n\log n} \right)$, can serve as an upper bound for distributed information spreading time \footnote{In contrast, the gossip time on the complete graph, $\Theta\left( {\log n} \right)$, may serve as a lower bound.}. Note that other meaningful performance benchmarks may also be used in our following discussion.

\subsubsection{Velocity Constrained Mobility} In this case, we are interested in determining how much velocity can make up for the deficiency in network connectivity due to small transmission range.
We mainly consider the $v_{\max} > r$ case\footnote{Intuitively, if node mobility is further limited by the transmission range, one would not expect that it can compensate for the network connectivity deficiency efficiently. In such scenarios, mobility may still be helpful for information spreading, which will be further explored in our future work.}. By comparing the result of \eqref{general-velocity-conduct-simplified} with the static conductance of a ring graph $\Theta (1/n)$, we can obtain the following velocity threshold for effective information spreading:
%
\begin{equation}\label{effective-velocity}
v_{\max}^{th} = \Omega \left( \frac{1}{n^2 r^2} \right).
\end{equation}
\emph{Remarks:} Since $v_{\max}=O(1)$ according to our model, a further analysis of \eqref{effective-velocity} reveals the following interesting points. (1) When $r=\Omega(1/n)$, the above velocity is what is needed to compensate for the connectivity deficiency so that the same information spreading performance is achieved as in a worst-case connected graph. In this case, there exists a tradeoff between $v_{\max}$ and $r$, in which the \emph{effective velocity} is inversely proportional to the square of transmission range $r$. (2) When $r=O(1/n)$, even fully random mobility cannot recover the spreading time of $\Theta \left( {n\log n} \right)$. Nonetheless, information can still be spread to the whole network given sufficiently high velocity, only at a slower speed (see discussion in III.A).

%
%
%
%
%

\subsubsection{Partially Random Mobility} Similarly, by comparing the result of \eqref{general-partial-conduct-simplified} with the static conductance of a ring graph $\Theta (1/n)$
a necessary condition for achieving the same information spreading performance as in a worst-case connected graph can be obtained as
\begin{equation}
{k(2n-k)}{r^2} = \Omega \left( {1} \right).
\end{equation}

\emph{Remarks:} There exists a tradeoff between the mobility ratio $k/n$ and $r$ for effective message spreading. With the reasonable assumption that $k/n=o(1)$\footnote{Otherwise the network is effectively fully random.}, the \emph{effective mobility ratio} is given by
\begin{equation}
\left({\frac k n}\right)^{th} = \Omega \left( \frac{1}{n^2 r^2} \right)=\omega (1/n).
\end{equation}

\subsubsection{One-Dimensional Mobility} Following the same approach as above,
%
a necessary condition for effective message spreading with this mobility model is
\begin{equation}
{{n_V}{n_H}}{r^2} = \Omega \left( {1} \right).
\end{equation}

\emph{Remarks:} There exists a tradeoff between the mobility balance $\frac{{{n_V}{n_H}}}{n^2}$ and $r$ for effective message spreading. Given $r$, the \emph{effective mobility balance} is given by
\begin{equation}
\left(\frac{{{n_V}{n_H}}}{n^2}\right)^{th} = \Omega \left( \frac{1}{n^2 r^2} \right).
\end{equation}
Given that $n_V+n_H=n$, we may further infer that both $n_V$ and $n_H$ should at least remain growing with $n$ when $r=o(1/\sqrt{n})$.

\section{Conclusions and Future Work}
\noindent This study, together with our previous work \cite{Mobile-conductance}, presents a unified analytical framework for information spreading in mobile networks that can accommodate various types of mobility patterns and different combinations of transmission range and moving speed. One future direction is to explore a multi-step move-and-gossip model to deepen our study in this area.


\vspace{0.25in}

\bibliographystyle{unsrt}

\begin{thebibliography}{}

\end{thebibliography}


\begin{thebibliography}{100}

\bibitem{Gossip}
D. Shah, ``Gossip algorithms," \emph{Foundations and trends in
networking}, vol. 3, no. 1, pp. 1-125, April 2008.

\bibitem{ZKongMobiHoc2008}
Z. Kong and E. Yeh, ``On the latency for information dissemination in mobile wireless networks,"
in \emph{Proc. ACM MobiHoc}, 2008, pp. 139-148.


\bibitem{Sunlei2013CRBlackhole}
L. Sun and W. Wang, ``Understanding Blackholes in large-scale Cognitive Radio Networks under Generic Failures," in \emph{Proc. IEEE INFOCOM}, Turin, Italy, 2013, pp. 728-736.

\bibitem{Xiaoming2014ComST}
X. Chen, H. Chen, and W. Meng, ``Cooperative Communications for Cognitive Radio Networks-from Theory to Applications," to appear in \emph{IEEE Communications Surveys \& Tutorials}, 2014.

\bibitem{Sunlei2013MobileCRN}
L. Sun and W. Wang, ``On distribution and limits of information Dissemination Latency and Speed in Mobile Cognitive Radio Networks," in \emph{Proc. IEEE INFOCOM}, Shanghai, China, 2011, pp. 246-250.


\bibitem{Spreading-MEG}
A. Clementi, A. Monti, F. Pasquale, and R. Silvestri, ``Information spreading in stationary markovian evolving graphs,"  \emph{IEEE Trans. Parallel Distrib. Syst.}, vol. 22, no. 9, pp. 1425-1432, Sept. 2011.

\bibitem{AClementiICALP2009}
A. Clementi, F. Pasquale, and R. Silvestri, ``{MANETS}: {H}igh mobility can make up for low transmission power," in \emph{Proc. ICALP}, 2009, pp. 387-398.

\bibitem{PJacquetTIT2010}
P. Jacquet, B. Mans, and G. Rodolakis, ``Information propagation speed in mobile and delay tolerant networks," \emph{IEEE Trans. Inf. Theory}, vol. 56, no. 10, pp. 5001-5015, Oct. 2010.

\bibitem{Pettarin2011}
A. Pettarin, A. Pietracaprina, G. Pucci, and E. Upfal, ``Tight bounds on information dissemination in sparse mobile networks," in \emph{Proc. PODC}, 2011, pp. 355-362.

\bibitem{Gossip-mobility}
A. Sarwate and A. Dimakis, ``The impact of mobility on gossip
algorithms,"  \emph{IEEE Trans. Inf. Theory}, vol.58, no.3, pp. 1731-1742, Mar. 2012.

\bibitem{Mobile-conductance}
H. Zhang, Z. Zhang and H. Dai, ``Mobile Conductance and Gossip-based Information Spreading in Mobile Networks," in \emph{Proc. IEEE ISIT}, 2013, pp. 824-828.

\bibitem{Mobile-conductance-TWC}
H. Zhang, Z. Zhang and H. Dai, ``Gossip-Based Information Spreading in Mobile Networks," \emph{IEEE Transactions on Wireless Communications}, vol. 12, no. 11, pp. 5918-5928, April 2013.




\bibitem{Mobility-throughput}
M. Grossglauser and D. Tse, ``Mobility increases the capacity of ad
hoc wireless networks," \emph{IEEE/ACM Trans. Networking}, vol. 10,
no. 4, pp. 477-486, Aug. 2002.

\bibitem{Li2013GC}
C. Li and H. Dai, ``Connectivity of Multi-channel Wireless Networks under Jamming Attacks," in \emph{Proc. IEEE GLOBECOM}, Atlanta, GA, 2013.

\bibitem{Velocity-mobility}
Y. Chen, S. Shakkottai, and J. G. Andrews, ``Sharing multiple messages over mobile networks," in \emph{Proc. IEEE INFOCOM}, 2011, pp. 658-666.

\bibitem{One-dim-mobility-1}
S. Diggavi, M. Grossglausser, and D. Tse,  ``Even one-dimensional mobility increases the capacity of wireless networks," \emph{IEEE Trans. Inf. Theory}, vol. 51, no. 11, pp. 3947-3954, Nov. 2005.


\bibitem{Mobile-conductance-general-R}
H. Zhang, H. Dai, Z. Zhang and Y. Huang, ``Mobile conductance under general transmission radius $r$," Department of Electrical Engineering, NC State University, Tech. Rep., 2013, Available at \url{http://www4.ncsu.edu/~hdai/InformationSpreading-HZ-TP2.pdf}.





\end{thebibliography}


\newpage

\end{document}